\documentclass[a4paper]{jpconf}
\usepackage{graphicx}
\usepackage{mathrsfs}
\usepackage[percent]{overpic}
\usepackage{amsmath}
\usepackage{setspace}
\usepackage[margin=2.5cm]{geometry}
\begin{document}
\title{Large-scale configuration interaction description of the structure of nuclei  around $^{100}$Sn and $^{208}$Pb}

\author{Chong Qi}

\address{Department of Physics, KTH Royal Institute of Technology, SE-10691 Stockholm, Sweden}

\ead{chongq@kth.se; https://www.kth.se/profile/chongq/}

\begin{abstract}
In this contribution I would like to discuss briefly the recent developments of the nuclear configuration interaction shell model approach. As examples, we apply the model to calculate the structure and decay properties of low-lying states in neutron-deficient nuclei  around $^{100}$Sn and $^{208}$Pb that are of great experimental and theoretical interests.
\end{abstract}

\vspace{-0.5cm}
\section{Introduction}
The FAIR-NUSTAR facility aims at addressing fundamental nuclear physics questions including: How are complex nuclei built from their basic
constituents? What are the limits for existence of nuclei? How to explain collective phenomena from
individual motion?  Nuclear theory plays a critical role in explaining those emerging phenomena and new data from challenging measurements on radioactive beam facilities as well as in predicting the structure and decay properties of nuclei in regimes that are not accessible in the laboratory. 
Nuclear theory also has mutually beneficial interplay with other many-body physics including atomic physics and condensed matter physics as well as astrophysics and cosmology.
The so-called \textit{ab initio} approaches (in the sense that realistic nucleon-nucleon interaction are applied without much \textit{ad hoc} adjustment), the nuclear shell model defined in a finite model space and the density functional theory 
are among the most commonly employed nuclear models that have been developed. Since all audience are not familiar with nuclear structure theory, firstly I would like to give an introduction to it.

\begin{figure}[htp]
\vspace{-0.5cm}
\begin{center}
\includegraphics[scale=0.34]{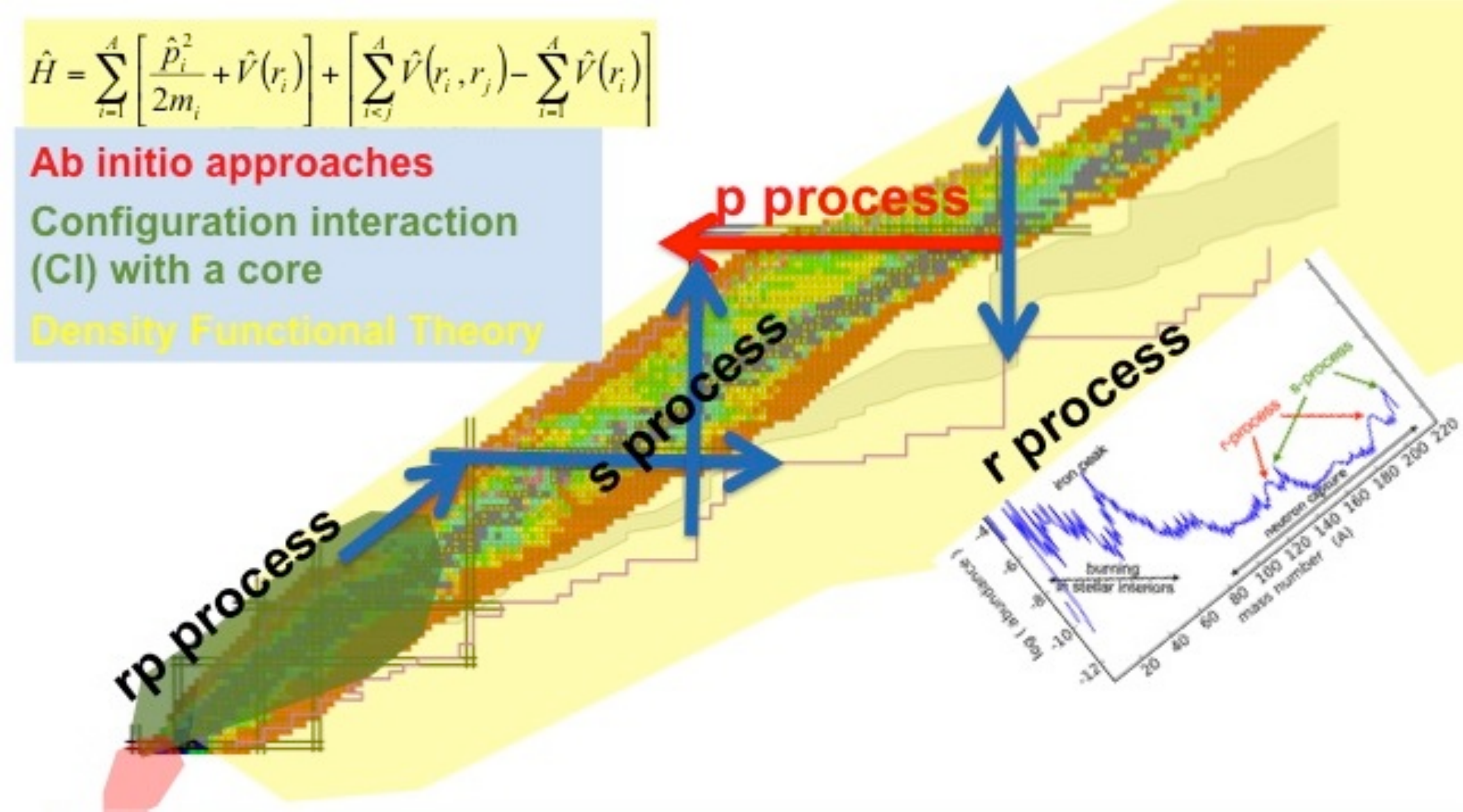}\includegraphics[scale=0.225]{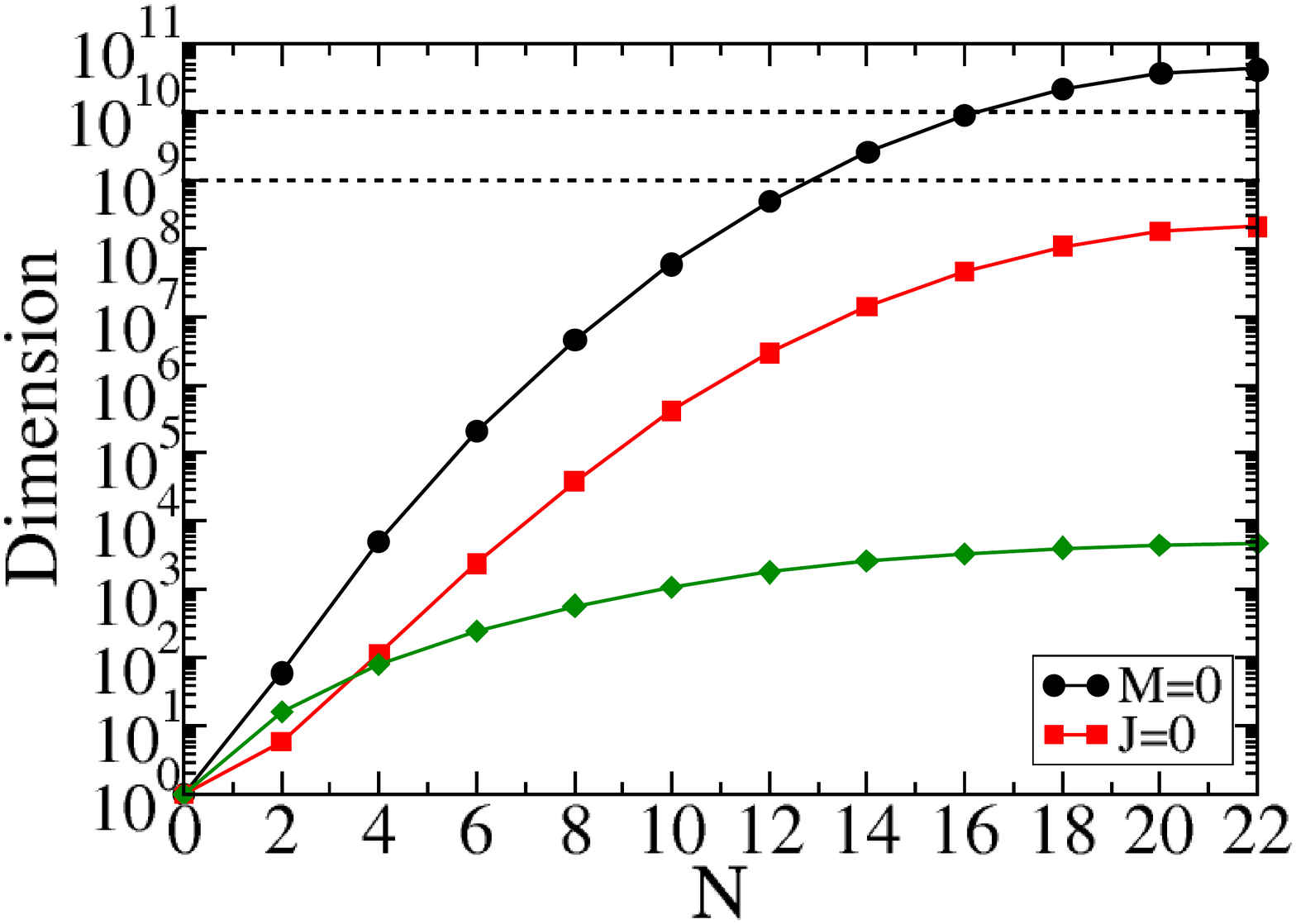}
\end{center}
\vspace{-0.7cm}
\caption{Left: Schematic plot for the territories of different nuclear models. Heavier nuclei around the arrows can also be described within the shell model approach. Right: Numbers of $M^{\pi}=0^+$ and $I^{\pi}=0^+$ shell-model states in even-even Pb isotopes as a function of valence neutron numbers $N$ in the model space defined by orbitals $2p_{1/2,3/2}$, $1f_{5/2,7/2}$, $0i_{13/2}$, $0h_{9/2}$.  \label{dimension}}
\end{figure}

The guiding principle for microscopic nuclear theory is that the building blocks of the nucleus, protons and neutrons, can be approximately treated as independent particles moving in a mean field that represents the average interaction between all particles.
The single-particle motion  provides a zeroth-order picture of the nucleus on top of which one has to consider the residual interaction between different particles. 
Within the \textit{ab initio} family, the no-core shell model approach aims at considering the residual correlation between all nuclei in a large space defined by the harmonic oscillator. As a result, only light nuclei below $^{16}$O can be evaluated (see, Fig. \ref{dimension}). 
The nuclear shell model, as we call it, is a full configuration interaction approach. It considers the mixing effect of all possible configurations within a given model space.
The model space is usually defined by taking a few single-particle orbitals near the Fermi surface. The number of orbitals one can include is highly restricted due to computation limitation. As an example, the dimension for the Pb isotopes are given in the right panel of Fig. \ref{dimension}. Despite of this challenge, the nuclear shell model is by far the most accurate and precise theory available on the market. 
State-of-the-art configuration interaction algorithms are able to diagonalize matrices with dimension up to $2\times10^{10}$ ($\sim10^{9}$ with the inclusion of three-body interaction or if only identical particles are considered). 

Below I will give a brief review on the challenges and recent developments of the nuclear configuration interaction shell model approach at our group. I will also mention a few simple but efficient truncation schemes. Shell-model calculations have been shown to be very successful in describing nuclei below $^{100}$Sn. Now we aim at giving a microscopic description of heavier nuclei, in particular those around the shell closures marked by arrows in Fig. \ref{dimension}, which may be possible with the help of Petaflops supercomputer. I will show some results we obtained for intermediate-mass and heavy nuclei around
$^{100}$Sn and $^{208}$Pb. 
I will explain the structure and decay studies of those nuclei, regarding both experimental and theoretical opportunities.
It may be interesting to mention that a proper description of the $N=126$ isotones
is important for our understanding of the astrophysical r process and its third peak.

\section{Configuration interaction shell model and the effective interaction}
The residual interaction between valence particles around the Fermi surface is mostly supposed to be of two-body nature. 
A common practice is to express the effective Hamiltonians in terms of single-particle energies and two-body matrix elements  as
\begin{eqnarray}
H_{eff}&=&\sum_{\alpha}\varepsilon_{\alpha}{\hat N}_{\alpha} + \frac{1}{4}\sum_{\alpha\beta\delta\gamma JT}\langle {\alpha}{\beta}|V|{\gamma}{\delta}\rangle_{JT}A^{\dag}_{JT;{\alpha}{\beta}}A_{JT;{\delta}{\gamma}},
\end{eqnarray}
where we have assumed isospin symmetry in the effective Hamiltonian, $\alpha=\{nljt\}$ denote the single-particle orbitals and $\varepsilon_{\alpha}$ stand for the corresponding single-particle energies. $\hat{N}_{\alpha}=\sum_{j_z,t_z}a_{\alpha,j_z,t_z}^{\dag}a_{\alpha,j_z,t_z}$ is the particle number operator. $\langle {\alpha}{\beta}|V|{\gamma}{\delta}\rangle_{JT}$ are the two-body matrix elements coupled to spin $J$ and isospin $T$. $A_{JT}$ ($A_{JT}^{\dag}$) is the fermion pair annihilation (creation) operator.  The two-body matrix elements can be calculated from realistic nucleon-nucleon potential where one has to consider the effect of its short range repulsion and the core polarization effects induced by the assumed inert core. Moreover, an 
optimization of the monopole interaction is necessary in most cases due to the neglect of three-body and other effects.
The total energy of the state $i$ is calculated to be
\begin{equation}
E^{\rm tot}_i =C+N\varepsilon_{0}+\frac{N(N-1)}{2}V_1+[T(T+1)-\frac{3}{4}N]V_0+ E_i^{\rm SM},
\end{equation}
where the constant $C$ denotes the (negative) binding energy of the core and $ E_i^{\rm SM}$ is the shell model energy. $\varepsilon_{0}$ is a mean single-particle energy.
The relative value of the $T=0$ and $T=1$ monopole interaction $V_{0,1}$  determines the relative position of the nuclear states with different total isospin $T$.
One may rewrite the Hamiltonian as $H_{eff}=H_m+H_M$ where $H_m$ and $H_M$ denote the (diagonal) monopole and Multipole Hamiltonians, respectively. The shell model energies can be written as
\begin{eqnarray}
\nonumber E_i^{\rm SM}
 &=& \sum_{\alpha}\varepsilon_{\alpha}<\hat{N}_{\alpha}>+\sum_{\alpha\leq\beta}V_{m;{\alpha}\beta}\left<\frac{\hat{N}_{\alpha}(\hat{N}_{\beta}-\delta_{\alpha\beta})}{1+\delta_{\alpha\beta}}\right>+\langle\Psi_i|H_M|\Psi_i\rangle,
\end{eqnarray}
where $\sum_{\alpha}<\hat{N}_{\alpha}> =N$, $\Psi_i$ is the calculated shell-model wave function of the state $i$.

Truncations often have to be applied in order to reduce the size of the shell-model bases. 
The simplest way of truncation is to restrict the maximal/minimal numbers of particles in different orbitals. This method is applied both to no-core and empirical shell model calculations.
In Ref. \cite{Back2013} we studied the structure and electromagnetic transition properties of light Sn isotopes within the large  $gdsh_{11/2}$ model space by restricting the maximal number of four neutrons that can be excited out of the $g_{9/2}$ orbital.
However, the convergence can be very slow if there is no clear shell or subshell closure or if single-particle structure are significantly modified by the monopole interaction, as it happens in neutron-rich light nuclei (see, e.g., Ref. \cite{Xu2013247}).

For systems involving the same kind of particles, the low-lying states can be well described within the seniority scheme \cite{tal93}. This is related to the fact that the $T=1$ two-body matrix elements in Eq. (1) is dominated by monopole pairing interactions with $J=0$. The seniority quantum number is related  to the number of particles that are not paired to $J=0$. Our recent studies on the seniority coupling scheme may be found in Refs. \cite{PhysRevC.82.014304,Qi2011a,Qi2012a,Qi2012g}. One can also derive the exact solution of the pairing Hamiltonian by diagolizing the matrix spanned by the seniority $v=0$, spin $I=0$ states which represent only a tiny part of the total wave function. This is applied in Ref. \cite{Xu2013247,PhysRevC.92.051304, Changizi2015210}.  The seniority coupling will be broken if both protons and neutrons are present where neutron-proton (np) coupling may be favored instead.
There has been a long quest for the possible existence of np pairing in $N\sim Z$ nuclei for which there is still  no conclusive evidence after long and extensive  studies (see, recent discussions in Refs. \cite{Ced11,Frauendorf201424,Qi11,Xu2012,Qi2015}).

One can evaluate the importance of a given basis vector $\psi_{i}$ within a partition through a perturbation measure
 \begin{equation}
 \label{Eg}
R_i = \frac{|\langle \psi_{i} | H_{eff} | \psi_{c} \rangle| }{\epsilon_{i} - \epsilon_{c}}
\end{equation}
where $\psi_{c}$ is the chosen reference   with unperturbed  energy $\epsilon_{c}$.
It is expected that the basis vectors with larger $R_i$ should play larger role in the given state dominated by the reference basis $\psi_{c}$, from which truncation scheme can be defined. 
 The off-diagonal matrix elements $\langle\psi_{i}|H_{eff}|\psi_{c}\rangle$ are relatively weak in comparison to the diagonal ones. The most important configurations may be selected by considering the difference of unperturbed energy difference as
$
r_i =  \epsilon_{i} - \epsilon_{c}.$
An truncated model space can thus be defined by taking those with smallest $r_i$.  The challenge here is that the truncated bases may not conserve angular momentum.
An angular momentum conserved correlated basis  truncation approach is introduced in Ref. \cite{PhysRevC.90.024306}.
Alternatively, one may consider an importance truncation based on the total monopole energy as \cite{Qi2016t}
\begin{eqnarray}
\nonumber E^{\rm m}_P&=& \sum_{\alpha}\varepsilon_{\alpha}{N}_{P;\alpha}+\sum_{\alpha\leq\beta}V_{m;{\alpha}\beta}\frac{N_{P;\alpha}(N_{P;\beta}-\delta_{\alpha\beta})}{1+\delta_{\alpha\beta}},
\end{eqnarray}
where $N_{P;\alpha}$ denotes the particle distributions within a given partition $P$. One can order all partitions according to the monopole energy $E^{\rm m}_P$ and consider the lowest ones for a given truncation calculation. 
The idea behind is that the Hamiltonian is dominated by the diagonal monopole channel. The monopole interaction can change significantly the (effective) mean field and drive the evolution of the shell structure.

\section{Selected results}
We have been evaluating the structure and decay properties of nuclei following the arrows as indicated in Fig. \ref{dimension}.
The robustness of the $N=Z=50$ shell closures has been studied extensively recently, which has fundamental influence  on our understanding of the structure of nuclei around the presumed doubly magic nucleus $^{100}$Sn. 
It was argued that $^{100}$Sn may be a soft core in analogy to the soft $N=Z=28$ core $^{56}$Ni. It seems such a possibility can be ruled out based on indirect information from recent measurements in this region \cite{Back2013,PhysRevC.84.041306,10.1038/nature11116,PhysRevLett.110.172501,PhysRevC.91.061304}. It is still difficult to measure the single-particle states outside the $^{100}$Sn core. The neutron single-particle states $d_{5/2}$ and $g_{7/2}$ orbitals  in $^{101}$Sn are very close to each other. A flip between the $g_{7/2}$  and $d_{5/2}$ orbitals from $^{103}$Sn to $^{101}$Sn was suggested in Ref. \cite{PhysRevLett.105.162502}.
This result was used in the construction of the effective Hamiltonian \cite{PhysRevC.86.044323} where the monopole interaction was optimized by fitting to all low-lying states in Sn isotopes using a global optimization method. The effective interaction has been shown to be successful in explaining many properties of nuclei in this region. The asymmetric electric quadrupole (E2) transition shape in Sn isotopes is suggested to be induced by the Pauli blocking effect \cite{Back2013}. A systematic study on the E2 transition in Te isotopes is done in Ref.~\cite{PhysRevC.91.061304}.

\begin{figure}[htp]
\vspace{-0.7cm}
\begin{center}
\includegraphics[scale=0.29]{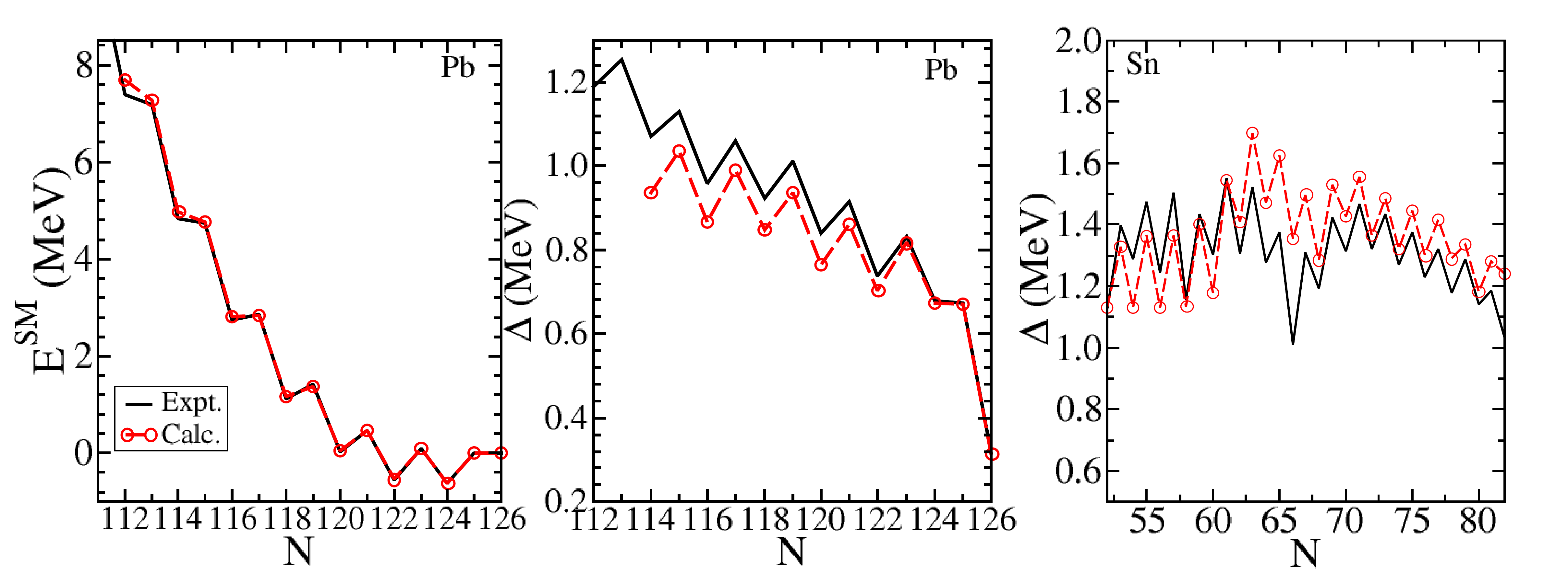}
\end{center}
\vspace{-0.7cm}
\caption{Shell-model energies for Pb isotopes with neutron number $N<126$ (left) and neutron odd-even mass staggering (empirical pairing gaps) in Pb (middle) and Sn (right) isotopes extracted from the experimental and calculated binding energies. \label{delta}}
\end{figure}

We have done truncation calculations  for two isotopes $^{200,194}$Pb by considering the relative importance as defined by the monopole Hamiltonian and monopole+diagonal pairing Hamiltonian. Convergence can be reached with a small portion (around 10\%) of the total M-scheme wave function in both cases.
We have also done pair-truncated shell-model calculations with collective pairs as building blocks in Refs. \cite{Xu2012,Qi2016t,PhysRevC.88.044332} for both the standard shell model and continuum shell model in the complex energy plane.
One example is the proton-unbound nucleus $^{109}$I \cite{Procter2011118} for which the level structure and E2 transition properties are very similar to those of
$^{108}$Te \cite{PhysRevC.84.041306} and $^{109}$Te \cite{PhysRevC.86.034308}, indicating that the odd proton in $^{109}$I is  weakly coupled to the $^{108}$Te daughter nucleus like a spectator.

In Fig.~\ref{delta} we plotted the calculated shell-model energy for Pb isotopes and compared them with experimental data.
Those energies are defined in the hole-hole channel relative to the assumed core $^{208}$Pb. 
For nuclei heavier than $^{196}$Pb, the difference between theory and experiment is less than 100 keV. The largest deviation appears in the case of $^{194}$Pb for which the calculation overestimate $E^{SM}$ by 300 keV. The empirical pairing gaps can 
be extracted from the binding energy by using the simple three-point formula, which carry important information on the two-nucleon pair clustering as well as $\alpha$ clustering in the nuclei involved \cite{Andreyev2013,Qi2014203}. 
 In nuclear systems the pairing collectivity
manifests itself through the coherent contribution of many shell-model configurations, which lead to large pairing gaps.
The results for Pb and Sn isotopes
are shown as a function of the neutron number in the middle and right panels of Fig. \ref{delta}. The overall agreement between experiments and calculations on the pairing gaps are quite satisfactory. Noticeable differences are only seen for mid-shell nuclei $^{196-198}$Pb and mid-shell Sn isotopes.

Our shell-model calculations can reproduce well the excitation energies of the low-lying $0^+$ and $2^+$ states in isotopes $^{198-206}$Pb. The excitation energies of the first $2^+$ isotopes show a rather weak parabolic behavior In the lighter Pb isotopes the excitation energy of the second $0^+$ state decreases rapidly with decreasing
neutron number. It even becomes the first excited state in $^{184-194}$Pb. 
Within a shell-model context, those low-lying $0^+$ states may be interpreted as coexisting deformed states which are induced by proton pair excitations across the $Z=82$ shell gap. The energy of those core-excited configurations get more favored in mid-shell Pb isotopes in relation to the stronger neutron-proton correlation in those nuclei. 

Another interesting phenomena is the nearly linear behavior of quadrupole moments in Cd, Sn as well as Pb isotopes for states involving the $h_{11/2}$ and $i_{13/2}$ orbitals, which can be explained in terms of shell occupancy within the seniority coupling scheme.
 As the occupancy increases, the quadrupole moments follow a linear decreasing trend and eventually vanish around half-filling.
 
There has already been a long effort answering the question whether the formation probabilities of neutron-deficient $N\sim Z$ isotopes  are larger compared to those of other nuclei \cite{Seweryniak2006,Liddick2006}. 
The $\alpha$ decays from $N\sim Z$ nuclei can provide an ideal test ground for our understanding of the np correlation.
We have evaluated within the shell-model approach the nn and pp two-body clustering in $^{102}$Sn and $^{102}$Te and then evaluated the correlation angle between the two pair by switching on and off the np correlation \cite{Qi2015}. If the np correlation is switched on, in particular if a large number of levels is included, there is significant enhancement of the four-body clustering at zero angle. This is eventually proportional to the $\alpha$ formation probability. It should be mentioned that, one need large number of orbitals already in heavy nuclei in order to reproduce properly the $\alpha$ clustering at the surface. The inclusion of np correlation will make the problem even more challenging due to the huge dimension.

\section{Summary}

We present briefly our recent works on the configuration interaction shell model calculations. A simple truncation scheme can be established by considering configurations with lowest monopole energies. Good convergence for Pb isotopes is reached for both the energy and wave function. Large scale calculations are carried out to study the spectroscopic and transition properties of nuclei around $^{100}$Sn and $^{208}$Pb that cannot be reached by standard shell model calculations. Both the ground state binding energies and excitation energies of low-lying states of the Sn and Pb isotopes can be reproduced very well.

\section*{Acknowledgement}
This work is supported by the Swedish Research
Council (VR) under grant Nos. 621-2012-3805, and
621-2013-4323. The
computations were partly performed on resources
provided by the Swedish National Infrastructure for Computing (SNIC)
at PDC, KTH, Stockholm.

\section*{References}
%

\providecommand{\newblock}{}


\end{document}